# Melting of spin ice state and development of fifth order susceptibility with magnetic field in pyrochlore $Tb_2Sn_2O_7$

Karan Singh, Dheeraj Ranaut, G. Sharma and K. Mukherjee

School of Basic Sciences, Indian Institute of Technology Mandi, Mandi 175005, Himachal Pradesh, India

**Abstract**

Pyrochlores offer an ideal playground to investigate the magnetic ground state of frustrated magnetic systems. In this class of materials, competition between various magnetic interactions remains frustrated and prevents an ordered magnetic state at low temperatures. $Tb_2Sn_2O_7$ has recently attracted significant attention due to its ordered spin-ice state. Additionally, in such systems, application of external magnetic field might result in exotic magnetic states. Our current investigation on $Tb_2Sn_2O_7$ reveal the presence of a new phase associated with fifth order susceptibility at low temperatures and high magnetic fields. In this compound, at zero fields, for a stabilized spin-ice state, the singlet-singlet state separated by δ play an imperative role. Under magnetic fields, δ increases and the Zeeman energy associated with the magnetic anisotropy is believed to get enhanced; which can be the key ingredient for evolution of higher-order moments, above 10 kOe, in this compound.

## 1. Introduction

In pyrochlores $R_2M_2O_7$ (R = rare earth and M = transition or sp metal), geometrical arrangements of the magnetic ions lead to magnetic frustrations [1, 2]. This family of compounds provides a platform to investigate the exotic phases which are of current interest in condensed matter physics, like, cooperative paramagnetic state [3], topological spin glass [4, 5], quantum spin liquids [6, 7], quantum spin ice [8-10]. Crystalline electric field (CEF) effect, geometrical frustration associated with nearest neighbor exchange interaction and dipole-dipole interaction act as the controlling parameter of physical properties of various phases in pyrochlores. Crystal field breaks the degeneracy of the magnetic ion multiplet down to Kramer doublets or a mixture of non-Kramer doublets and singlets [11]. As a result, it leads to the quantum effects along with the development of higher-order moments, for example, dipole-octupolar doublets [12], quadrupolar moments [11, 13]. Both Kramer and non-Kramer ground states can carry the higher-order moments associated with the magnetic dipole moments [12-13]. Evidence of quantum effects along with the higher-order magnetization has been recently reported in spin-ice related Kramer and non-Kramer pyrochlores [14-16]. In this context, $Tb_2Sn_2O_7$ (a non-Kramer pyrochlore), is another promising candidate. The controversy regarding the ground state of this compound is not yet settled despite using various experimental and theoretical tools. Initially in $Tb_2Sn_2O_7$, the ground state was reported to be ferromagnetic below $T_C$ ~ 0.87 K [17]. Further, other investigations revealed the presence of an ordered spin ice state at $T_C$, followed by strong fluctuations below $T_C$ [18-23]. A partially ordered state associated with the two-sublattice ferrimagnetic long-range ordering was also reported for the same compound [24]. J. Zhang *et al* had discussed "soft" spin ice ordering at 0.87 K, where the doublet ground and excited state are separated by an energy $\Delta$ ~ 17.4 K (~ 1.5 meV) [25]. However, the study reported by Y. Chapuis *et al* highlighted that there is a non-degenerate ground state followed by an excited single level at energy of $\delta$ ~ 2 K (i.e., singlet-singlet state) and a doublet at a energy of ~ 17 K [26]. Later, it was debated that the excited singlet level is found to be much lower than the value 2 K [27]. Thus, from above literature reports, it can be inferred that local tetragonal distortion breaks the degeneracy of doublet ground state resulting in an excited single level of energy $\delta$, which may be responsible for stabilizing the spin ice ordered state in $Tb_2Sn_2O_7$. H. R. Molavian *et al*., theoretically addressed the quantum ordered state boundary in pyrochlore, particularly, focusing on how the ground state is tuned by the perturbation of exchange and dipole-dipole interaction [28]. Similarly, in $Tb_2Sn_2O_7$, it is expected that presence of nearest-neighbor exchange interaction, dipole-dipole interaction, and magnetic anisotropy may compete among themselves with variation in temperature and magnetic field and enter a new phase of matter, associated with exotic physical properties. Muon spin relaxation studies on $Tb_2Sn_2O_7$, in the presence of magnetic fields reveal that, low-energy magnetic density of states of the collective excitations increase with magnetic fields [29]. Further, through the investigations of collective excitations, higher-order moments have been studied in other candidates [30-31]. In most circumstances, such moments are also signaled by the change in bulk properties such as the heat capacity [32]. Although there are several reports on $Tb_2Sn_2O_7$, but to our best knowledge there were no attempts made to cover the following queries: whether with magnetic fields i) ordered spin-ice state is stable; ii) crystal field splitting energy is modified; iii) there is any phase transitions associated with higher-order moments.

In this manuscript, we have addressed these queries through magnetization, heat capacity, and non-linear DC susceptibility studies on $Tb_2Sn_2O_7$. Our results reveal that the splitting δ between the singlet-singlet states is enhanced in the presence of magnetic field, which may possibly destabilize the ordered spin-ice state. On increasing the magnetic field above 10 kOe, Zeeman energy related with the magnetic anisotropy starts dominating, which results in the formation of magnetic phase associated with fifth order susceptibility. To further support our observations, we have proposed a theoretical model to calculate δ as a function of applied magnetic field. Our investigations indicate towards an increment of δ on application of magnetic field, which stabilizes the interactions among higher-order moments.

## 2. Experimental details

Polycrystalline $Tb_2Sn_2O_7$ was synthesized by the conventional solid state reaction method. $Tb_2O_3$ (Sigma Aldrich, ≥99.99% purity) and $SnO_2$ (Sigma Aldrich, ≥99.999% purity) with an appropriate molar ratio were mixed and grinded. The mixed powder was kept in the alumina crucible at $1280^0C$ in ambient atmospheric conditions for 36 hours thrice, with intermediate grinding. The obtained powder was then finally pelletized and sintered at $1280^0C$ for 36 hours. Powder x-ray diffraction (XRD) measurement was carried out at 300 K using Rigaku Smart Lab instruments with CuK*α* radiation. Rietveld refinement of the x-ray diffraction data was performed using FullProf Suite software. Temperature (*T*) and magnetic field (*H*) dependent DC and AC magnetization were performed using Magnetic Property Measurement System (MPMS) from Quantum design, USA. Heat capacity measurements were performed using Physical Property Measurement System (PPMS) from Quantum design, USA. Raman spectra was collected at room temperature in backscattering geometry by using Horiba HR Evolution spectrometer with 532 nm excitation laser under low laser power (7 mW) to avoid local heating. X-ray photoelectron spectra were obtained using a monochromatic AlKα (1486.6 eV) X-ray source with an energy resolution of 400 meV and Scienta analyser (R3000). The sample surface was cleaned in situ by scraping with a diamond file until sample has the minimum O 1s signal as measured by AlKα x-ray. The binding energy was calibrated by measuring the Fermi energy of Ag in electrical contact with the sample. The base pressure during the measurement was 5× $10^{-10}$ mbar.

## 3. RESULTS

### 3.1 Crystal structure

Rietveld refinement of the XRD pattern obtained at 300 K shows that $Tb_2Sn_2O_7$ crystallizes in the cubic structure with space group $Fd\bar{3}m$ and is in single phase (figure 1). This structure has two independent anion sites O (1) and O (2), occupying the 48f and 8b position, respectively. 8a position, i.e., O (3), remains vacant. Larger Tb cation occupies the 16d position coordinated to six O (1) atoms and two O (2) atoms, while smaller Sn cation occupies the 16c position which is coordinated to six O (1) atoms at equal distance. The additional reflections of (331), (551) planes are an indication of stabilized pyrochlore structure [33]. The pyrochlore structure is also reflected in the Raman spectra taken at room temperature (see text and figure S1 of supplementary information). In addition, we have performed X-ray photoemission spectroscopy (XPS) measurements at 300 K which indicate the presence of $Tb^{3+}$ and $Sn^{4+}$ valence state of $Tb_2Sn_2O_7$ (see supplementary information for detail and figure S2).

### 3.2 Magnetization

Figure 2 (a) shows the temperature dependent DC susceptibility (*M/H*) at different fields in the temperature range 1.8 to 30 K. At 0.1 kOe, it is observed that magnetic susceptibility increases with decreasing temperature and no signature of magnetic ordering is seen up to 1.8 K. Under application of magnetic fields, a change of slope in susceptibility appears around 2.5 K for 10 kOe (shown by arrow). At 30 kOe, susceptibility seems to saturate below 6.6 K (inset of figure 2 (a)). Above 30 kOe, saturation in susceptibility appears below 9.8 K and 12 K for 50 kOe and 70 kOe, respectively. This change in the behavior of susceptibility indicates that magnetic field alters the exchange interactions among magnetic moments. Curie-Weiss (CW) law at 0.05 kOe is fitted using the equation:

$$M/H = C/(T-\theta) \quad \ldots\ldots\ldots (1)$$

where $C$ is the Curie constant and $\theta$ is the CW temperature. Figure 2 (b) illustrates that above 25 K, inverse susceptibility curve is fitted well with CW law. Below this temperature, a non-linear deviation is observed which could be due to CEF effect. The obtained values of C and $\theta$ are ~ 11.70 emu/mol-Oe-K and $\theta$ ~ -11.47 K, respectively. These values are in good agreement with previously reported values [17, 20]. The experimentally obtained effective magnetic moments $\mu_{eff}$ (= $2.8\sqrt{C}$ $\mu_B$ ~ $9.58\mu_B$), are almost close to the theoretically calculated $\mu_{eff}$ for $Tb^{3+}$ [calculated using relation: $\mu_{eff} = g\sqrt{J(J+1)}$ $\mu_B$, where $g = 1.5$, $S = 3$, $L = 3$, $J = 6$]. The negative value of $\theta$ indicates the dominance of antiferromagnetic interactions in this compound. Figure 2 (c) shows the magnetization (*M*) as a function of magnetic field at 1.8 K. It is observed that magnetization increases with increasing magnetic field and reaches around 5 $\mu_B$/f. u. at 1.8 K and 70 kOe which match with the previous reports [18, 20]. In addition, it is noted that a change in slope occurs at about 10 kOe (inset of figure 2 (c)). With increasing temperature, this slope change feature is weakened resulting in its suppression above 6 K (figure S3 of supplementary figure). In order to get a further insight, we have measured the temperature dependent AC susceptibility. The real part of AC susceptibility ($\chi'$) in the field range of 0-20 kOe is shown in the figure 3. At 0 Oe, $\chi'$ increases with cooling and no change of slope is observed up to 1.8 K. In the presence of magnetic fields, a maximum around 2.3 K (at 3 kOe) is noted. Above this field, the maximum widens and shifts toward higher temperature (shown by arrow in the inset of figure 3). The field response of the temperature, $T_{max}$ (temperature corresponding to maxima in $\chi'$) shows a linear behavior. The straight-line extrapolation of these points (as $T_{max}$) gives a value 0.85 K at 0 Oe (as shown in Figure S4 of supplementary information). Interestingly, this value is same where the ordered spin-ice transition has been reported for $Tb_2Sn_2O_7$ [18, 19, 21, and 22]. The shift of this spin-ice transition to higher temperature with magnetic field (< 3 kOe) is also reported below 1.8 K [21]. Here, we would like to mention that we have not observed a significant signal of imaginary part ($\chi''$) in the measured range of temperatures and magnetic fields (not shown). Further from the inset of figure 3, it is noted that the observed broad maximum is suppressed around 12 kOe. This shifting of the maximum to higher temperatures on application of magnetic field may be due to the melting of spin-ice state, and it may give rise to the development of a new phase. This evolution of new phase is clearly visible in the derivative of magnetization with respect to field (inset of Figure 2 (c)), where a clear change in slope is observed around 10kOe. As reported in Ref [18], ordering at 0.85 K results from the effective ferromagnetic (FM) interaction among Ising spins, which is a consequence of the collective effect of antiferromagnetic (AFM) nearest neighbor exchange interactions $J_{nn}$ along <110> edge axes of the tetrahedron, FM nearest neighbor dipole-dipole interaction $D_{nn}$ along <100> axes and Zeeman energy in combination with magnetic anisotropy $D_a$ along and perpendicular to local <111> direction. This would be possible when $D_{nn}$ dominates $J_{nn}$. The development of the new phase at high fields (as mentioned earlier) may be caused either by the dominance of $J_{nn}$ or $D_a$. When $J_{nn}$ dominates, ground state becomes extremely frustrated; and a spin liquid like phase develops [18]. This is in contradiction to our case as our temperature dependent magnetization data shows the saturation at low temperatures and high fields. So, it might be possible that $D_a$ dominate over $J_{nn}$ at high fields. In order to substantiate the preceding state, we have measured the temperature dependent heat capacity in the presence of magnetic fields.

### 3.3 Heat capacity

Figure 4 shows the temperature dependence of heat capacity (*C*) at different fields in the temperature range 1.8-15 K. In this temperature range, it is noted that lattice contribution is negligible (illustrated in figure S5 and the corresponding text in the supplementary information). At 0 Oe, below 15 K, heat capacity increases with decreasing temperature and a hump appear with a maximum around temperature 5.2 K. The value of maxima is around 3.6 J/mol-K; a similar value has been reported at the same temperature for $TbSn_2O_7$ in Ref [26]. It is assigned to the Schottky anomaly.

Various crystal-field level schemes are proposed for this Schottky anomaly in the Refs [24, 26, and 27]. In most reports, it is believed that this system has a singlet-singlet state with energy difference δ and excited doublet state at energy Δ ~ 17 K [26, 27]. With this energy level scheme, we have fitted the Schottky anomaly of our heat capacity data for 0 kOe (see text and inset of figure S5 in the supplementary information). In Ref [9], it is stated that entropy (*S*) increases and reaches the value of Rln2 at around 4.4 K. At this temperature, the thermal energy ($k_BT$) is smaller than Δ; however, it is larger than the δ [26]. In general, for a two-level system with the same degeneracy of ground and excited energy levels, *S* is around Rln2. In this case, both levels are almost equally populated when thermal energy becomes equal to energy between ground and excited levels. Hence, it can be said that, in $Tb_2Sn_2O_7$, $S$ has reached the value of Rln2 due to equally populated ground state energy levels separated with δ. Above 4.4 K, entropy further increases due to populations of the higher-lying CEF levels [24].

Further, as clearly visible from the figure 4, with increasing magnetic fields up to 10 kOe, Schottky temperature shows a downward shift, and the height of the maxima increases. Above 30 kOe, it is noted that temperature of maxima in heat capacity shifts to higher temperature which is believed to give rise to a new distinct state, different from the low-field ground state. It can be seen that the entropy (derived from the integration of *C/T*) in low and high field regime is significantly different from each other (figure S6 in supplementary information). From the fitted heat capacity data with the proposed CEF levels scheme, it is concluded that δ increases with the magnetic fields. The increment in δ reduces the frustrations which may be responsible for shifting of spin-ice transition to higher temperatures. Thus, the weak hump observed in heat capacity around 2.4 K for 5 kOe (shown by arrows in the inset of Figure 4(a)) may be attributed to the shifting of spin-ice transition with field. Similar kind of behavior is also observed in our χ' measurements. In the high field regime (> 10 kOe), it is suspected that Zeeman energy associated with the magnetic anisotropy $\mathcal{D}_a$ play a noteworthy role as discussed in the magnetization section. This anisotropy energy may be associated with the evolution of higher-order moments [9, 13, 29]. To study this phase, we use the non-linear susceptibility technique.

**3.4 Non-linear DC susceptibility**

Non-linear DC susceptibility is an important technique to detect the presence of higher-order spin correlations. Investigation of higher order moment fluctuation helps to probe the evolution of new phase. In general terms, magnetization in the direction of applied magnetic field can be expanded as [34-36]:

$$M/H = \chi_1 + \chi_3 H^2 + \chi_5 H^4 \ldots\ldots (3)$$

where $\chi_1$ is the linear term associated with dipolar moments, $\chi_3$ and $\chi_5$ are the non-linear terms associated with higher-order moments. For determination of $\chi_3$ and $\chi_5$, magnetization data (under field cooled condition) was collected with slow cooling rate of 5 mK/s at a constant field, similar to the protocol as mentioned in Ref [34, 35]. Figure 5 (a) shows *M/H* plotted as a function of $H^2$.

The figure shows that *M/H* follows a quadratic path with $H^2$ above 10 kOe and then increases rapidly with a further decrease in magnetic field. With the increasing temperature, the sharp increase of *M/H* and slope of the quadratic path decreases. In the field range 10-65 kOe, a good quadratic fit to the experimental data *M/H* (as a function of $H^2$) is obtained (For representational purpose, the fitted curve at 1.8 K is shown in the inset of figure 5 (a)). The extracted parameters $\chi_1$, $\chi_3$ and $\chi_5$ are plotted as a function of temperature. The temperature dependent $\chi_1$ is found to be in analogy with that observed directly from DC magnetization measurements (as shown in figure S7 of supplementary information). Figure 5 (b) and (c) represents $\chi_3$ and $\chi_5$ as a function of temperature, respectively. It is observed that $\chi_3$ is negative over the entire measured temperature range. This behavior of $\chi_3$ may be due to negative curvature of the Brillouin function in finite fields [36]. In case of $\chi_5$, it is found to be positive at 1.8 K. It decreases linearly with increasing temperature and a crossover from positive to negative can be obtained at 13 K by linear extrapolation (red curve in figure 5 (c)). The observed positive behavior of $\chi_5$ can be understood on the basis that, two lowest CEF states ($E_1$ and $E_2$) of ground state separated by energy $\delta$ mix in the form $\frac{1}{\sqrt{2}}$ ( $|E_1\rangle \pm |E_2\rangle$) and render $Tb^{3+}$ as pseudospin-$\frac{1}{2}$ magnetic ion, which carries the composite object associated with the dipolar and higher-order moments [37, 38]. Interaction in such magnetic ions can be responsible for the positive $\chi_5$. For further studies, a two-level system with energy separation $\delta$ should be adequate to understand the evolution of new phase where $\chi_5$ becomes positive [39, 40]. With the assumption that magnetic field is parallel to the quantization axis, linear and non-linear terms of susceptibility can be described as [40]:

$$\chi_1 = \frac{\gamma^2}{\delta} \frac{1}{\tau} \frac{1}{1+A} \dots\dots\dots (4)$$

$$\chi_3 = \frac{\gamma^4}{3!\delta^3} \frac{1}{\tau^3} \frac{A-2}{(1+A)^2} \dots\dots\dots (5)$$

$$\chi_5 = \frac{\gamma^6}{5!\delta^5} \frac{1}{\tau^5} \frac{A^2-13A+16}{(1+A)^3} \dots\dots\dots (6)$$

where $A = 0.5\ e^{1/\tau}$ and $\tau = k_B T/\delta$. From the equation (5), it is found that $\chi_3$ is zero for $A = 2$ and $\chi_3 < 0$ for $A < 2$. As mentioned earlier, $\chi_3 < 0$ in the measured temperature range, implying that *A* should always be less than 2. Similarly, from the equation (6), it is calculated that $\chi_5 > 0$ for either $A < 1.38$ or $A > 11.63$. From above analysis, it can be inferred that *A* should be less than 1.38. Using the temperature (~ 13 K) where $\chi_5$ is expected to be 0, we estimate $\delta$. The value of $\delta$ is found to be ~ 13.11 K (1.13 meV). In the region of $\chi_5$, it is believed that the strength of exchange molecular field interacting with the magnetic field, along and perpendicular to local <111> direction is increased in the presence of magnetic fields [23]. As a result, Zeeman energy related with the magnetic anisotropy $\mathcal{D}_a$ becomes increasingly important and can be responsible for evolution of the new phase associated with positive fifth order susceptibility [9, 41, 42].

Furthermore, we study the magnetic free energy, to explore whether the field induced phase is stable or not. Magnetic free energy ($F$) can be written in term of $H$ and $M$ up to sixth order [40].

$$F = -HM + a_2 M^2 + a_4 M^4 + a_6 M^6 \quad \ldots\ldots (7)$$

where $a_2 = 1/2\chi_1$, $a_4 = -\chi_3/4\chi_1^4$, and $a_6 = [3\chi_3^2 - \chi_5\chi_1]/\chi_1^7$. It is noted that $a_2$ is positive, since $\chi_1 > 0$. While $a_4$ is also positive, since $\chi_3 < 0$ down to the lowest measured temperature 1.8 K. This suggests continuous phase transition. Inset of figure 5(c) shows the plot of temperature dependent $a_6$. It is found that it remains positive in the entire measured range of temperature. From this scenario, it can be stated that free energy is bounded and enter a stable phase transition which is associated with fifth order susceptibility.

## 4. Discussion

As mentioned in Section 3.1, $Tb_2Sn_2O_7$ crystallizes in cubic structure, in which $Tb^{3+}$ ion is surrounded by eight oxygen ions. The CEF created by these oxygen ions at the $Tb^{3+}$ site has $D_{3d}$ symmetry. The local symmetry at the $Tb^{3+}$ site can be described as a trigonal compression along one of its body diagonals, which is parallel to the local <111> directions. As a result, the CEF Hamiltonian acting on a $Tb^{3+}$ ion is invariant under three-fold rotation around the <111> direction and point inversion with respect to the $Tb^{3+}$ site. Based on these constraints, the CEF Hamiltonian ($H_{CEF}$) can be described as [25]

$$H_{CEF} = \alpha_J\, D_2^0 O_2^0 + \beta_J\, (D_4^0 O_4^0 + D_4^3 O_4^3) + \gamma_J\, (D_6^0 O_6^0 + D_6^3 O_6^3 + D_6^6 O_6^6) \quad \ldots\ldots (8)$$

where the $\alpha_J, \beta_J, \gamma_J$ are the coefficients and $O_n^m$ are the Stevens operator's equivalent. $D_n^m$ are the crystal field parameters which have been reported for $Tb_2Sn_2O_7$ in the Ref [25]. In the crystal structure of $Tb_2Sn_2O_7$, $Tb^{3+}$ ion has eight electrons in the 4$f$ shell with angular momentum $J = 6$. Under CEF, it splits into $2J + 1 = 13$ crystal field levels. In the lowest-energy sector, the ground state and the first excited state doublet is separated by energy $\Delta$. This separation energy is believed to determine the order of the magnetic moments. In order to understand the ground state interactions, we add the Hamiltonian which includes dipole-dipole interactions ($H_d$), and the

Hamiltonian of nearest neighbor exchange interaction ($H_e$) in the equation (9) for Ising spins [23]:

$$H = H_{CEF} + H_d + H_e \quad \ldots\ldots\ldots (9)$$

$$H_d = \mathcal{D}_{nn} \sum_{i>j} [J_i \cdot J_j - 3(J_i \cdot \hat{r}_{ij})(J_j \cdot \hat{r}_{ij})] \, |R_{ij}|^3 \quad \ldots\ldots (10)$$

$$H_e = - \mathcal{J}_{nn} \sum_{<i,j>} J_i \cdot J_j \quad \ldots\ldots (11)$$

where the $\mathcal{J}_{nn}$ (= $\mathcal{J}_{nn}^x, \mathcal{J}_{nn}^x, \mathcal{J}_{nn}^z$) are the nearest-neighbour exchange interaction terms and $R_{ij} = R_j - R_i = |R_{ij}|\hat{r}_{ij}$, where $R_i$ denotes the position of site $i$. Further, due to local tetragonal distortion in this compound, it is found that doublet ground state is splited with an energy $\delta$ resulting in a singlet-singlet ground state [26]. Taking into account the projection of the total angular momentum $J_i$ (=$J_i^x, J_i^y, J_i^z$) at the $i$-site in the ($xoz$) and ($yoz$) plane of the cube (where the z-axis is parallel to the <111> axes), the perturbed Hamiltonian ($H_{per}$) is added to the equation (9), [26]:

$$H = H_{CEF} + H_d + H_e + H_{per} \quad \ldots\ldots\ldots (12)$$

$H_{per}$ is expressed as:

$$H_{per} = (H_{per})_{xoz} + (H_{per})_{yoz} + H_m \quad \ldots\ldots\ldots (13)$$

$$(H_{per})_{xoz} = - \mathcal{D}_t \left[\frac{2}{3}(J_i^x)^2 + \frac{1}{3}(J_i^z)^2 + \frac{\sqrt{2}}{3}\overline{J_i^x J_i^z}\right] \quad \ldots\ldots (14)$$

$$(H_{per})_{yoz} = - \mathcal{D}_t \left[\frac{2}{3}(J_i^y)^2 + \frac{1}{3}(J_i^z)^2 + \frac{\sqrt{2}}{3}\overline{J_i^y J_i^z}\right] \quad \ldots\ldots (15)$$

where $\mathcal{D}_t > 0$ measures the scale of the crystal-field distortion from the cubic symmetry. The terms $\overline{J_i^x J_i^z}$ (=$J_i^x J_i^z + J_i^z J_i^x$), $\overline{J_i^y J_i^z}$ (=$J_i^y J_i^z + J_i^z J_i^y$) are associated with the pseudo-spin operators $\tau_i^x, \tau_i^y$, respectively [14] and $H_m$ accounts for the magnetic state below 0.85 K. When the magnetic field is applied, the Zeeman energy plays a significant leading role. It is supposed that z-component of the magnetic field ($H_z$) is coupled to $J_i^z$ via the $g_\parallel$ while x and y-components of the magnetic field ($H_x$ and $H_y$) are coupled to $\tau_i^x$ and $\tau_i^y$ via the $g_\perp$, respectively (where $g_\parallel$ and $g_\perp$ are the g-tensor components, which are along and perpendicular to the local <111> direction (considered along z-axis), respectively) [43]. The value of these tensors characterizes the anisotropy of the ground state. Irrespective of the orientation of the magnetic field, at zero or low values of the applied field, the ground state has a larger anisotropy associated with $g_\parallel \neq 0$. For large magnetic field, it is believed that the tensor $g_\perp$ becomes dominant leading to an enhanced separation of the singlet-singlet state. As a result, the wave-functions of the pseudo-spins can transfer the ground state in time-reversal even or time reversal odd higher-order moment [13, 14, and 44]. Hence, the equation (12) implies that the competition between Zeeman energy associated with magnetic anisotropy, dipole-dipole interactions, and exchange interactions $\mathcal{J}_{nn}$ (expressed in eqn (10) and (11)) can be responsible for developing distinct ground state under the application of magnetic fields.

In order to support our experimental observation, we have extended the theoretical model described in the Ref. 26, in order to calculate δ in the presence of external magnetic field. For this purpose, the doubly degenerate perturbation theory mentioned in Ref. 45 is used. This theory relates the perturbation Hamiltonian to the splitting induced by the perturbation,

$$k_B \delta = \left[ \left( \langle +|H_{per}|+\rangle - \langle -|H_{per}|-\rangle \right)^2 + 4\left| \langle -|H_{per}|+\rangle \right|^2 \right]^{1/2} \quad \ldots\ldots\ldots (16)$$

The doublet wave functions which characterize the unperturbed Hamiltonian are denoted by $|\pm\rangle$. The splitting δ between the singlet-singlet states corresponds to the difference between energy eigen values of doublet wave functions $|\pm\rangle$. According to Ref. 26, in the absence of magnetic field, the calculated value of δ has been found to be 2.5 K. Hence, by taking into account this preliminary information, the perturbed Hamiltonian (eqn. (13)) has been modified as:

$$H'_{per} = \begin{bmatrix} -\frac{1}{2}g_{||}H\mu_B \cos\theta + \frac{\delta}{2} & -\frac{1}{2}g_{\perp}H\mu_B \cos\theta\, e^{-i\phi} \\ -\frac{1}{2}g_{\perp}H\mu_B \cos\theta\, e^{i\phi} & \frac{1}{2}g_{||}H\mu_B \cos\theta - \frac{\delta}{2} \end{bmatrix} \quad \ldots\ldots\ldots (17)$$

Here, δ takes into account the splitting observed at zero field due to the terms mentioned in equ. (13) and the other terms in equ. (17) accounts for the Zeeman Hamiltonian. So, the difference between the energy eigenvalues ($E_1$ and $E_2$) corresponding to the above Hamiltonian gives the value of δ as a function of applied $H$ i. e. $\delta$ ($H$). One important thing to note here is that both anisotropic terms ($g_{||}$ and $g_{\perp}$) are considered in the Hamiltonian in order to observe the effect of polycrystallinity on the measured magnetization.

Hence, in order to find the value of δ ($H$), the values of $g_{||}$ and $g_{\perp}$ are needed which can be calculated by fitting the $M$ ($H$) data with a modified equation obtained from the equation (17). Average magnetization $\langle M \rangle$, can be calculated by considering a very general expression given as:

$$\langle M \rangle = \frac{m_1 e^{-E_1/k_B T} + m_2 e^{-E_2/k_B T}}{e^{-E_1/k_B T} + e^{-E_2/k_B T}} \quad \ldots\ldots\ldots (18)$$

Here, $m_1 = -\frac{\partial E_1}{\partial H}$ and $m_2 = -\frac{\partial E_2}{\partial H}$. $E_1$ and $E_2$ are the energy eigenvalues corresponding to the Hamiltonian described in the equ. (17). For the limiting case i.e., taking easy axis along $g_{||}$ (i.e., $g_\perp = 0$) and $\delta = 0$, the above equation reduces to the standard equation described in [46]:

$$\langle M \rangle = \frac{1}{2} g_{||} \mu_B \cos\theta \tanh\left(\frac{g_{||} H \mu_B \cos\theta}{2 k_B T}\right) \quad \ldots\ldots\ldots (19)$$

Further, by integrating equation (18) for all possible orientations, $\langle M \rangle$ can be calculated and the expression is given as:

$$\langle M \rangle = \frac{1}{4\pi} \iint_0^\pi \int_0^{2\pi} \frac{m_1 e^{-E_1/k_B T} + m_2 e^{-E_2/k_B T}}{e^{-E_1/k_B T} + e^{-E_2/k_B T}} \sin\theta \, d\theta \, d\phi \quad \ldots\ldots\ldots (20)$$

As also mentioned earlier, the $g_\perp$ tensor is expected to dominate at higher fields and thus it is appropriate to take $g_\perp$ as a function of magnetic field. Figure 6 (a) depicts the $M$ ($H$) curve at 2 K up to 70 kOe and red line shows the fitting of equation (20). By trying different $H$ dependence on $g_\perp$, we have concluded that ($g_\perp \sim H^{0.25}$) dependence shows the best match with the experimental data. The obtained values of the parameters $g_{||}$ and $g_\perp$ are ~ (12.23 ± 0.15) and ~ ((2.56 ± 0.05) × $H^{0.25}$) respectively. So, $g_\perp$ tensor increases with field and becomes more dominating at higher fields, which in turn is expected to increase the splitting of the singlet-singlet ground state at higher fields.

Now, $\delta$ ($H$) is calculated by integrating the energy difference ($E_2 - E_1$) over all possible orientations in order to compensate the powder-average and is given by:

$$k_B \, \delta \, (H) =$$

$$\frac{1}{4\pi} \iint_0^\pi \int_0^{2\pi} 2 \left( \frac{g_{||}^2 H^2 \mu_B^2}{2} \cos^2\theta + \frac{g_\perp^2 H^2 \mu_B^2}{2} \sin^2\theta + \frac{\delta^2}{4} - \frac{g_{||} H \mu_B}{2} \delta \sin\theta \right)^{1/2} \sin\theta \, d\theta \, d\phi \ldots (21)$$

Figure 6 (b) shows the deduced $\delta$ (in units of K) as a function of applied $H$. $\delta$ ($H$) follows a very interesting behavior remaining constant at low fields and then starts increasing linearly as a function of $H$. Thus, both from our experimental observations and theoretical model, it can be inferred that with increasing field, $\delta$ increases, resulting in the evolution of higher order moments above 10 kOe.

For $Tb_2Sn_2O_7$, the evolution of various ground states with magnetic fields is summarized through a $C/T$ color map in $H$-$T$ plane, as shown in figure (7). In this figure, three possible regions are extracted from magnetization, heat capacity and non-linear DC susceptibility measurements. Region I: This region is established from the temperatures where maximum in heat capacity ($C^{max}$) occur. In this region, highly degenerate ground state unveils where $\mathcal{J}_{nn}$ dominates for Ising spins and nature of region is paramagnetic. Region II: Here, the degeneracy is lifted, and it results in a singlet-singlet ground state and doublet excited state. The difference between energy levels of singlet-singlet ground state is δ where the ferromagnetic interaction associated with the angular momentum $J_i^z$ is dominating due to supremacy of dipole-dipole interactions $\mathcal{D}_{nn}$. It results in an ordered spin-ice phase, which is expected to be shifted in field range of 3-12 kOe, as determined from the maxima of $\chi_{ac}'$. In this region, an increment in magnetic field results in increment in δ. Region III: This region is ascribed by the development of fifth order susceptibility. As magnetic field increases, the separation between the singlet-singlet state, δ, is increased. As a result, above 10kOe, singlet-singlet state separated by δ mix, resulting in Tb acting as a pseudospin-$\frac{1}{2}$ magnetic ion. The fifth order susceptibility can be associated with interactions among pseudospin-$\frac{1}{2}$ magnetic ion [42, 47]. This area is demarcated from the saturation magnetization at low temperature and from non-linear susceptibility measurements. The crossover from ordered spin ice phase to higher ordered moment phase (around 12 kOe, where two lines meet) can be due to competition between $\mathcal{D}_{nn}$ and Zeeman energy associated with g-tensor. This result in higher ordered moments becoming more significant in this compound under the application of magnetic fields (above 12 kOe) due to enhancement in the Zeeman energy associated with $g_\perp$.

Finally, we would like to compare our system with the other conventional spin-ice systems. In this regard, we note a striking resemblance in the saturation magnetization value in ordered spin-ice system $Tb_2Sn_2O_7$ with that of conventional spin-ice systems, such as $Ho_2Ti_2O_7$ and $Dy_2Ti_2O_7$ [48, 49]. In zero or low magnetic field, it is found that the later system stabilizes the spin-ice state due to an effective ferromagnetic interaction which is associated with the dipole-dipole coupling. In high magnetic field, the spin-ice state is destabilized and a saturation in magnetization has been reported with an average moment of around 5 $\mu_B$/R (R = Dy and Ho) [50-52]. Similar value of the average moment has been found for $Tb_2Sn_2O_7$ [18, 20]. On the other side, in $Ho_2Ti_2O_7$ and $Dy_2Ti_2O_7$, the doublet ground and excited states are separated by energy Δ ~ 300-350 K and spin-ice state is associated with only the former state, following the ice-rule [48]. In contrast, in $Tb_2Sn_2O_7$, the doublet ground and excited states are separated by energy Δ ~ 17 K, which is very small in comparison to that of $Ho_2Ti_2O_7$ and $Dy_2Ti_2O_7$. As a result, the doublet-doublet picture is altered by the local tetragonal distortion, and it leads to the splitting of the doublet ground state into two singlets. Hence, in $Tb_2Sn_2O_7$, conventional spin-ice ordering is precluded. In high magnetic field, it is believed that this system acquires the higher-order moments. But up to our best knowledge, there are no report related to such moments in $Ho_2Ti_2O_7$ and $Dy_2Ti_2O_7$ although the ground state of all these acquires the almost same average saturation moments. We believe that our study might be beneficial to understand the other spin-ice pyrochlores carrying the higher-order moments.

## 5. Summary

In conclusion, application of magnetic field results in melting of ordered spin ice state and development of positive fifth order susceptibility, above 10 kOe, in $Tb_2Sn_2O_7$. Destabilization of spin ice state may be due to the increment in $\delta$. Above 10 kOe, due to increased $\delta$, Zeeman energy linked with magnetic anisotropy dominates over the exchange and dipole energy where higher order magnetic moments become considerable. Occurrence of quadrupole order has already been discussed in $Tb_2Ti_2O_7$ [11, 13], where the crystal field scheme is totally different from that of $Tb_2Sn_2O_7$. Our investigation may provide a pathway for researchers to explore the ground state in presence of magnetic field through microscopic experimental probes or through other theoretical studies.

**ACKNOWLEDGEMENTS**

The authors acknowledge IIT Mandi for experimental and financial support.

**ADDITIONAL INFORMATION**

Supplementary information

**Figures**

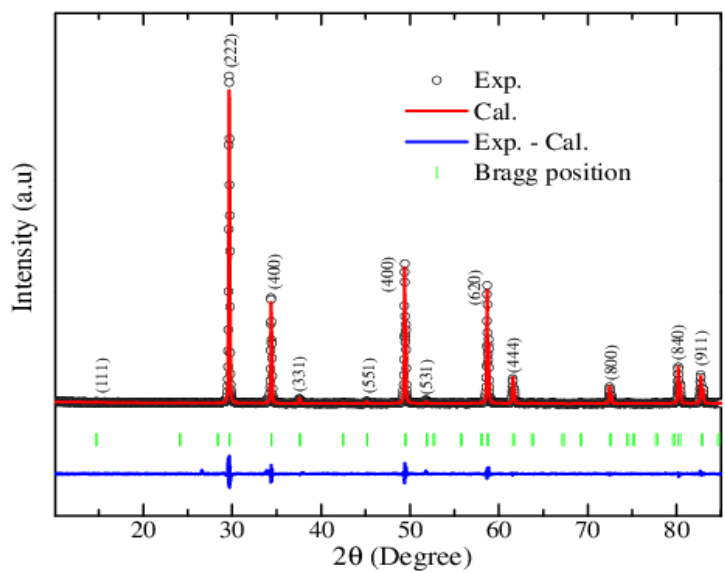

**Figure 1:** Rietveld refined experimental (open circles), calculated (red line) and difference (blue line) powder x-ray diffraction pattern at 300 K. The vertical bars correspond to the Bragg positions.

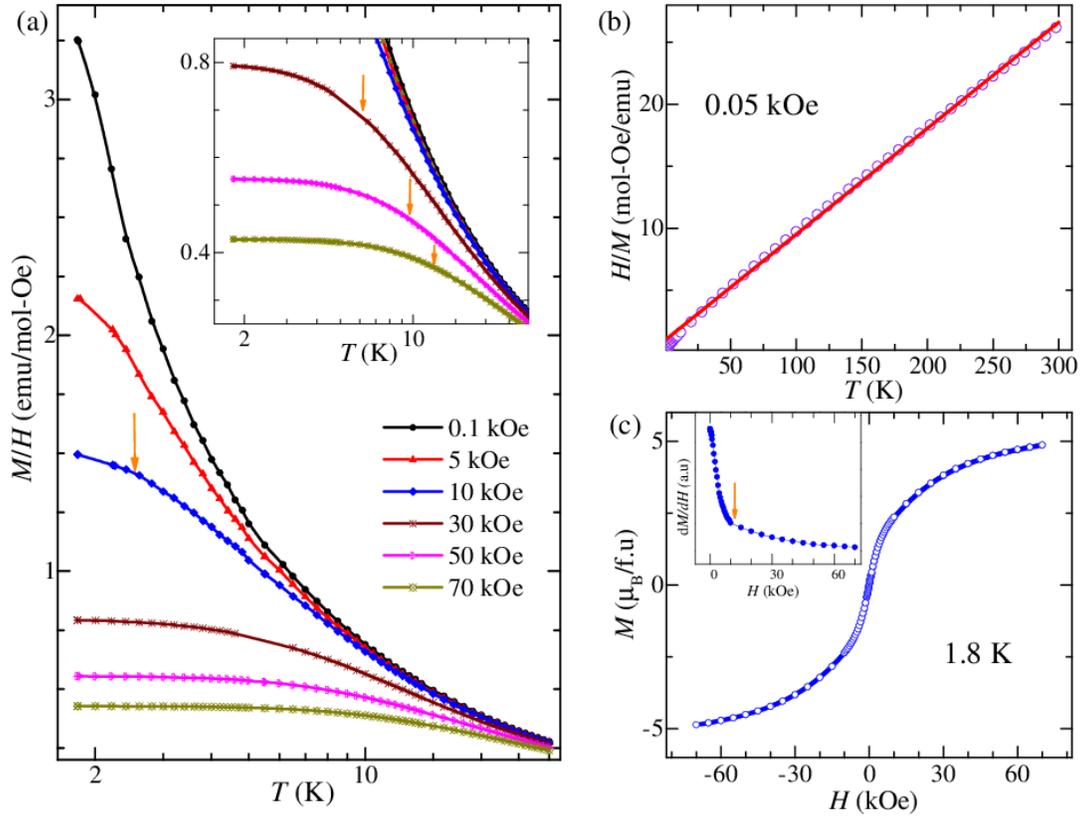

**Figure 2:** (a) Temperature dependent DC susceptibility (*M/H*) at different fields under zero field cooling condition. (b) Temperature dependent inverse DC susceptibility at 0.05 kOe. Red line denotes the Curie-Weiss law fitting. (c) Magnetization versus magnetic field curve at 1.8 K. Inset: Derivative of the magnetization with respect to field plotted as a function of magnetic field at 1.8 K. Arrow points to a change of slope around 10 kOe.

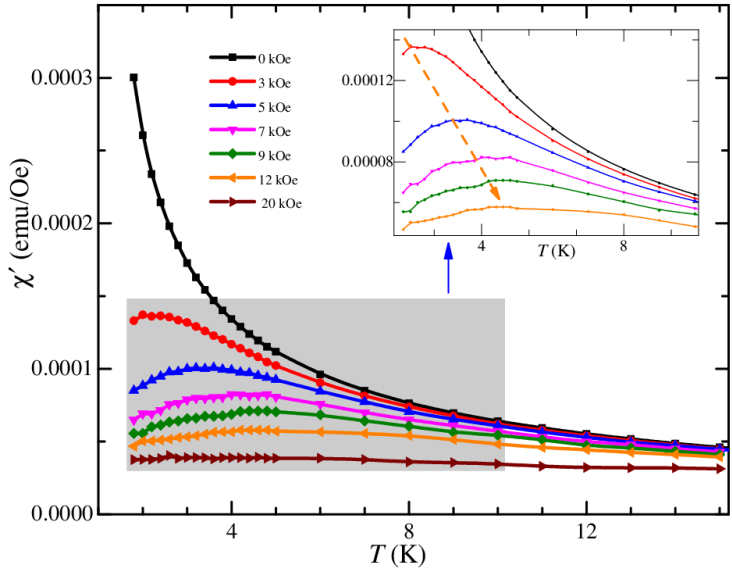

**Figure 3:** Temperature response of the real part of AC susceptibility (χ') (at 2 Oe, 30 Hz AC field) under various superimposed DC fields. Inset: Expanded view of the same figure where a maximum in AC susceptibility is observed.

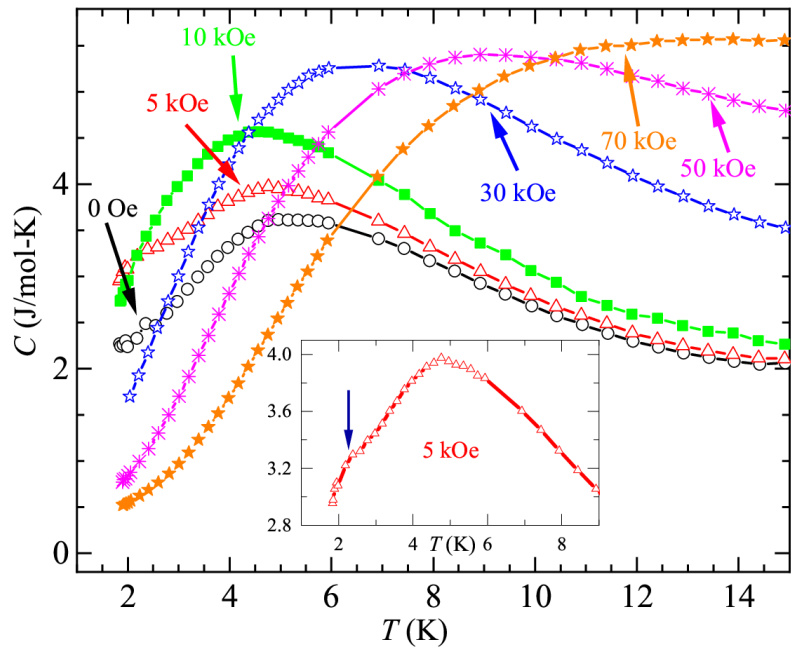

**Figure 4:** Temperature response of heat capacity at different fields. Inset: Expanded view of the heat capacity curve at 5 kOe.

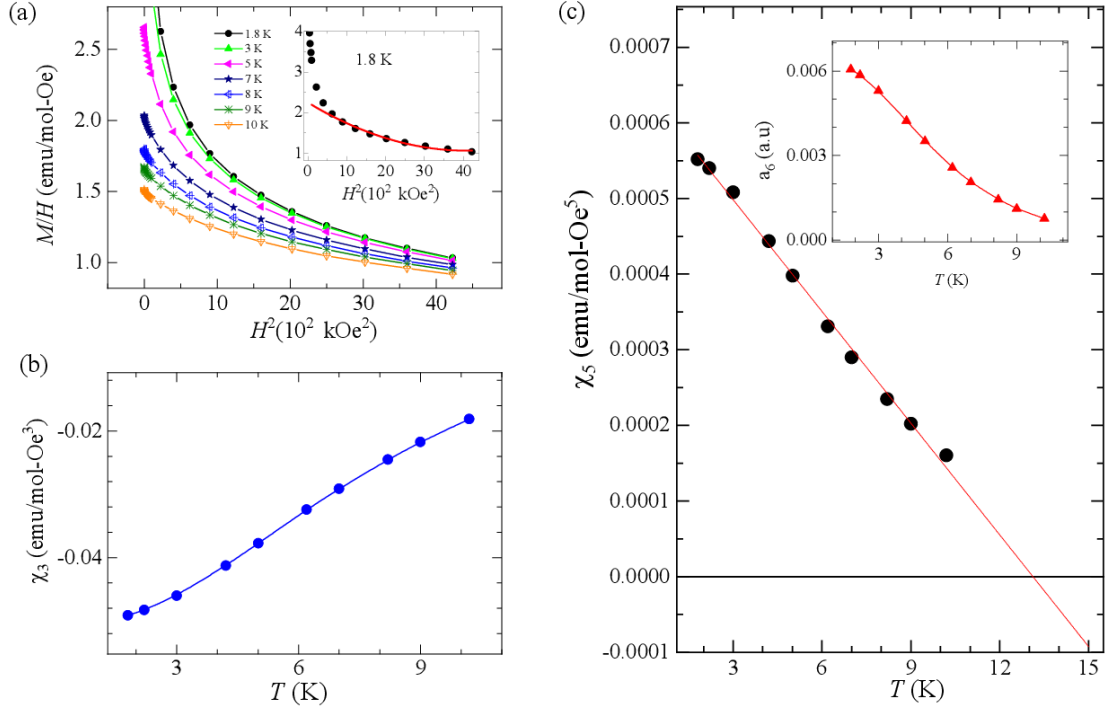

**Figure 5:** (a) *M/H* curves plotted as a function of $H^2$ at different temperature. The inset represents the same curve at 1.8 K with solid red line showing the fit to equation (3). (b) Temperature response of the 3$^{rd}$-order susceptibility. (c) Temperature response of the 5$^{th}$-order susceptibility. Inset: $a_6$ $\{= [3\chi_3^2 - \chi_5\chi_1]/\chi_1^7\}$ plotted as a function of temperature.

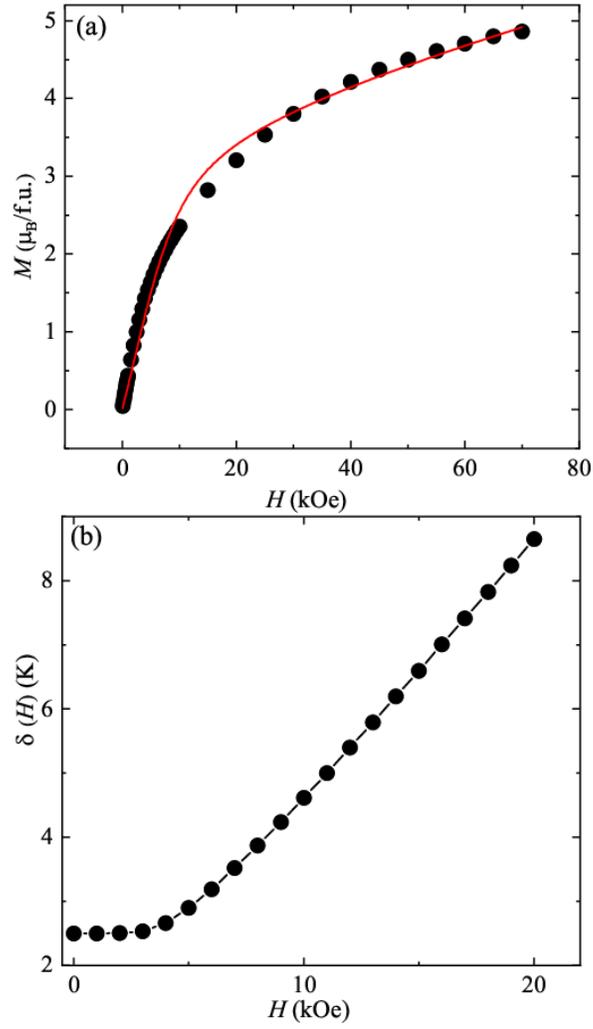

**Figure 6:** (a) *M* (*H*) curve obtained at 2 K. The red solid line shows the fitting of equation (20) (as described in text). (b) The splitting of the doublet ground state (δ) as a function of applied magnetic field (*H*) calculated by solving the equation (21) (as described in text).

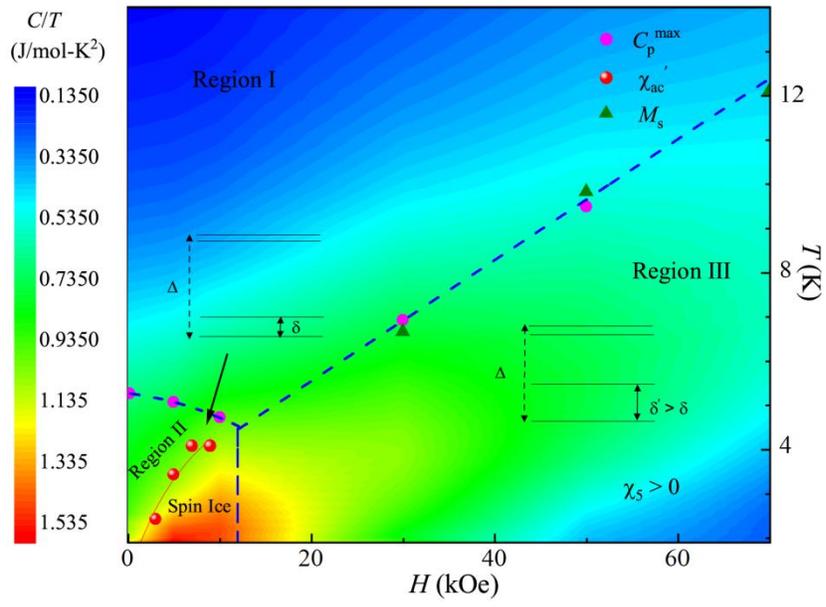

**Figure 7:** $C/T$ color map in $H$–$T$ plane for $Tb_2Sn_2O_7$.

# Supplementary information

# Melting of spin ice state and development of fifth order susceptibility with magnetic field in pyrochlore $Tb_2Sn_2O_7$

Karan Singh, Dheeraj Ranaut, G. Sharma and K. Mukherjee

School of Basic Sciences, Indian Institute of Technology Mandi, Mandi 175005, Himachal Pradesh, India

### Raman spectroscopy

Raman spectroscopy was carried out at 300 K (figure S1). We observed the Raman active modes associated to pyrochlore structure at ~ 310.83 $cm^{-1}$, 413.31 $cm^{-1}$, 503.83 $cm^{-1}$, and 529.57 $cm^{-1}$. These modes are assigned to one $A_{1g}$ (503.83 $cm^{-1}$), one $E_g$ (413.31 $cm^{-1}$) and two $T_{2g}$ (310.83 $cm^{-1}$, 529.57 $cm^{-1}$) [1]. These Raman modes correspond to the vibration of <Tb-O> and <Sn-O> bonds. Similar Raman spectra have been reported for pyrochlore $Tb_2Sn_2O_7$ [2].

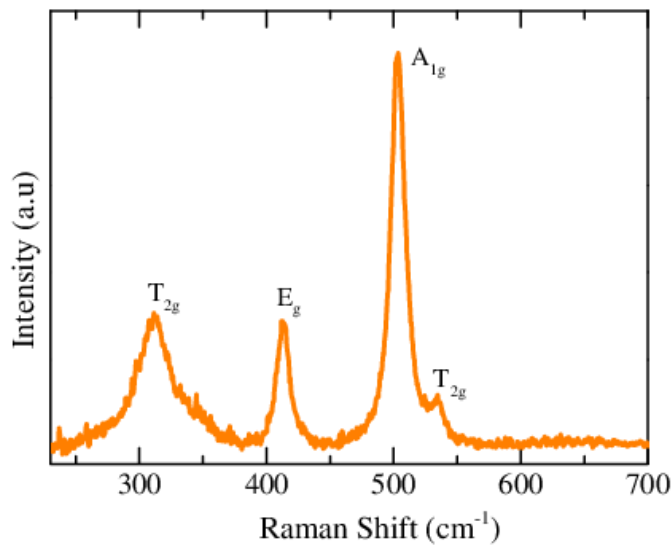

**Figure S1**: Room temperature Raman spectra.

**X-ray photoemission spectroscopy**

        To get an idea about the charge states of the constituent elements of $Tb_2Sn_2O_7$, X-ray photoemission spectroscopy (XPS) measurements at room temperature was performed to investigate the core level spectra of Tb-3$d$, Sn-3$d$, and O-1$s$. All spectra are fitted with the Casa software to extract the possible valence states in this compound. As per literature reports, for Tb-3$d$, two peaks are commonly seen and they are assigned to $3d_{5/2}$ and $3d_{3/2}$. We observed the major peaks at around 1240.4 eV and 1276.5 eV, indicating the presence of $Tb^{3+}$ ion [3]. A similar pattern of peaks was also reported in Ref [5]. In addition, we extract a very small peak at the higher binding energy from the fitting corresponding to the $Tb^{4+}$ ion. This could be due to presence of some surface impurities. The small peak at ~ 1249.27 eV is expected to be due to Tb satellite emission [4]. For Sn-3$d$, the peaks at ~ 486.3 eV and 494.70 eV are observed due to $3d_{5/2}$ and $3d_{3/2}$. It is determined that the observed binding energies are closest to that of $Sn^{4+}$ [5]. O-1$s$ core level spectrum shows a sharp peak at ~ 530.14 eV. The binding energy peak can be attributed to lattice oxygen [6]. Therefore, from the analysis of XPS data, it can be said that there is a presence of valence states of $Tb^{3+}$ and $Sn^{4+}$ in $Tb_2Sn_2O_7$ as a result these ions balance each other.

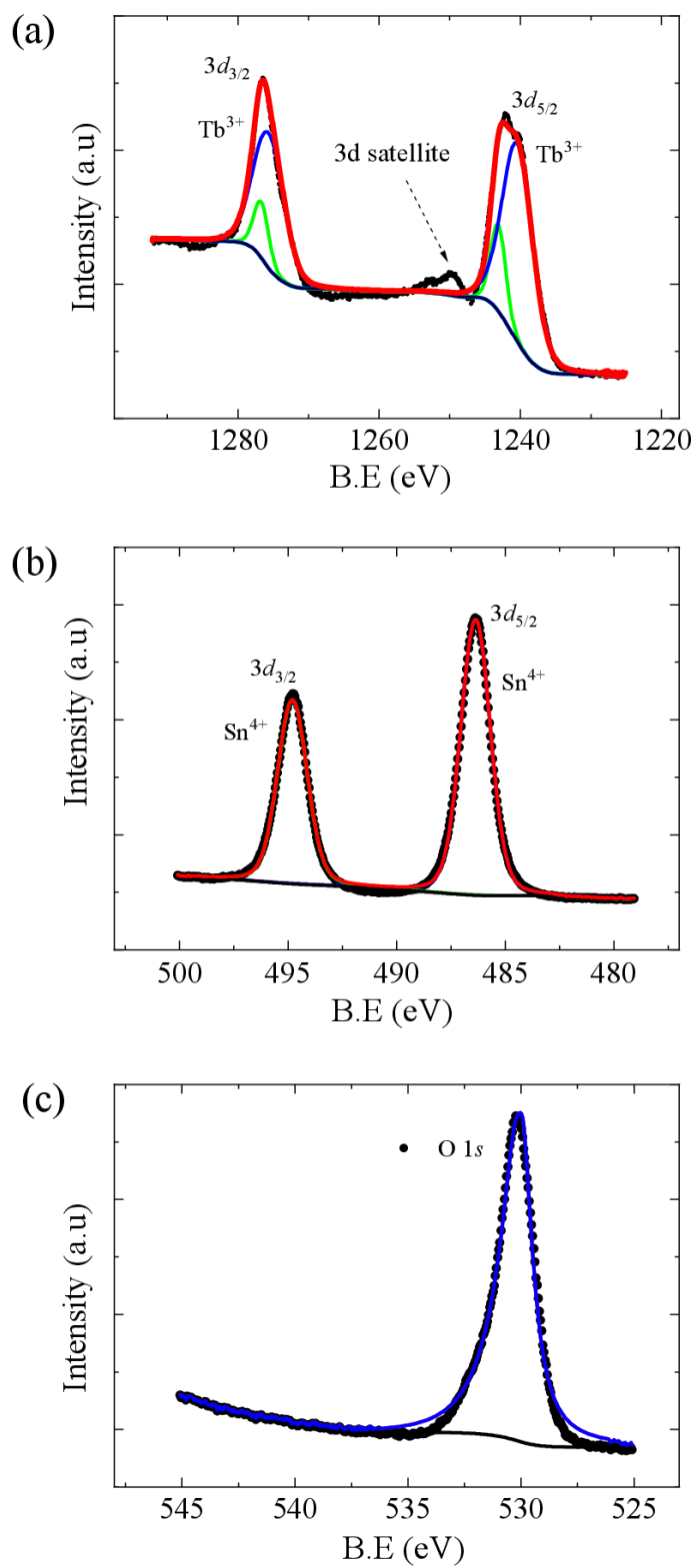

**Figure S2:** (a), (b), and (c) Room temperature energy-dispersive x-ray spectra of the core level of Tb $3d$, Sn $3d$, and O $1s$, respectively.

**Magnetism**

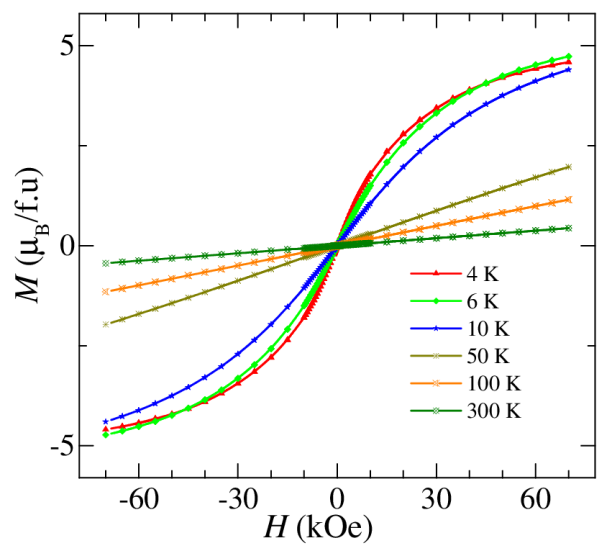

**Figure S3:** Magnetization as a function of magnetic field at different temperatures.

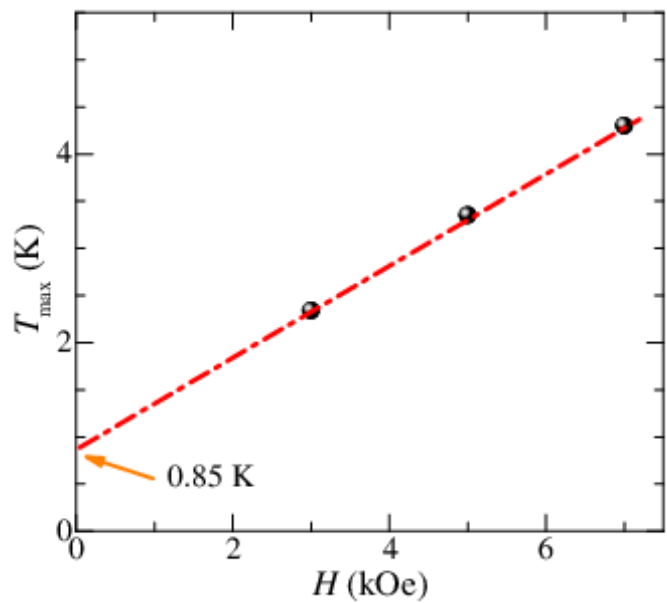

**Figure S4:** Magnetic field versus $T_{max}$ (corresponding to the maxima in the real part of AC susceptibility). Red line is the linear fitting through these points extrapolated to 0 Oe.

**Heat capacity**

Figure S5 shows the temperature dependence of heat capacity (*C*) at 0 Oe. Above 15 K, lattice part is assumed to have dominating contribution, which might be due to the presence of both acoustic and optical phonons. The optical phonons dominate at higher temperature region in comparison to the acoustic phonons. The contribution of both phonons is approximated with the Debye and Einstein models. The Debye model is associated with the acoustic phonons whereas Einstein model is associated with the optical phonons. To estimate the Debye and Einstein temperature, *C* is fitted using the equation [7]

$$C(T) = m\, C_{Debye}(T) + (1-m)\, C_{Einstein}(T) \quad\ldots\ldots\ldots (1)$$

where *m* is weightage of the Debye term. $C_{Debye}$ and $C_{Einstein}$ is the Debye and Einstein contribution to the lattice heat capacity, respectively, which can be described by equation given as [7].

$$C_{v\,Debye}(T) = 9nR \left(\frac{T}{\theta_D}\right)^3 \int_0^{\frac{\theta_D}{T}} \frac{x^4 e^x}{(e^x-1)^2} \quad\ldots\ldots (a1)$$

$$C_{v\,Einstein}(T) = 3nR \left(\frac{\theta_E}{T}\right)^2 \frac{e^{\frac{\theta_E}{T}}}{(e^{\frac{\theta_E}{T}}-1)^2} \quad\ldots\ldots (a2)$$

where n is the number of atoms per formula unit, R is universal gas constant, $\theta_D$ and $\theta_E$ are Debye temperature and Einstein temperature, respectively. A good fitting of the equation (1) is observed with the Debye temperature ~ 645 K and the Einstein temperature ~ 158 K, with *m* ~ 0.75 for Debye term (shown in red line in figure S5). From fitting, it can be seen that lattice contribution is negligible at low temperature, below 15 K. Hence, it can be said that heat capacity is dominated by electronic and magnetic contributions at low temperature.

Below 15 K, the Schottky anomaly of the heat capacity data is fitted to the crystalline electric field (CEF) levels scheme (shown in the inset of figure S5) for the singlet state followed by an excited single level at energy of δ (~ 2 K) and a doublet at a thermal energy of Δ ~ 17 K [8]. It can be written as [9]:

$$C = \frac{\beta^2(g_1 g_0 \delta^2 \exp(-\beta\delta) + g_0 g_2 \Delta^2 \exp(-\beta\Delta))}{(g_0 + g_1 \exp(-\beta\delta) + g_2 \exp(-\beta\Delta))^2} + \frac{g_1 g_2 \exp(-\beta(\delta+\Delta))[\delta(\delta-\Delta) + \Delta(\Delta-\delta)]}{(g_0 + g_1 \exp(-\beta\delta) + g_2 \exp(-\beta\Delta))^2} \quad\ldots\ldots (2)$$

where $g_0$, $g_1$ and $g_2$ are the degeneracy of the singlet ground state, first excited singlet state and second excited doublet state, respectively. $\beta = 1/kT$ ($k$ = Boltzmann constant).

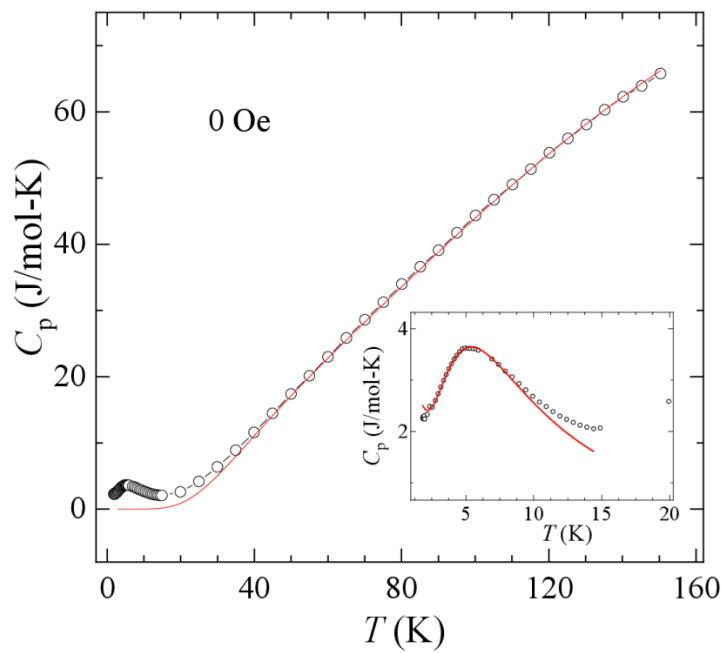

**Figure S5:** Temperature dependent heat capacity curve at 0 Oe. Red line indicates the fitting using equation (1). Inset shows the Schottky anomaly fitting to the equation (2).

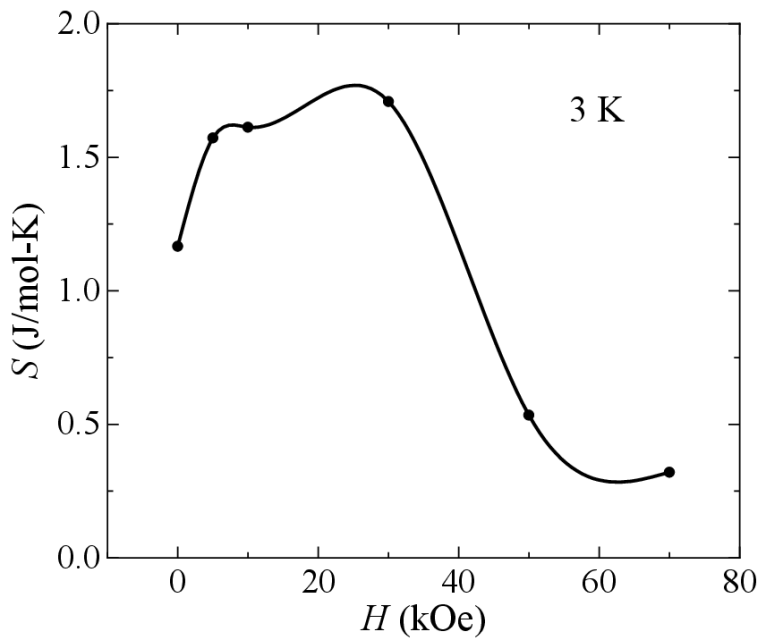

**Figure S6:** Magnetic field dependent entropy ($S$) at 3 K.

**Non-linear susceptibility**

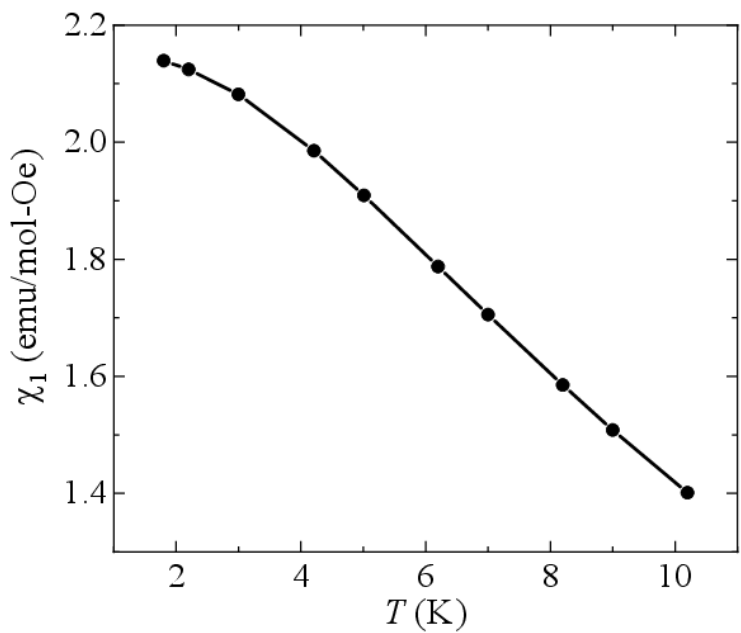

**Figure S7:** Temperature dependent $\chi_1$, extracted from non-linear susceptibility data.